%% file: main.tex
\documentclass[sigconf,9pt]{acmart}

\AtBeginDocument{%
  }

\usepackage{algorithmic}
\usepackage{lipsum} 
\usepackage{graphicx}
\usepackage{textcomp}
\usepackage{xcolor}
\usepackage{caption,xspace}
\usepackage{pifont}
\usepackage{enumitem}
\usepackage{textcomp,balance}
\usepackage{bm}
\usepackage{amsfonts}
\usepackage{url}
\usepackage{multirow}
\usepackage{booktabs}
\usepackage{longtable}
\usepackage{graphicx}
\usepackage{subcaption}
\usepackage{subfloat}
\usepackage{numprint}
\usepackage{ragged2e}
\usepackage{setspace}
\usepackage{framed}
\usepackage{tcolorbox}

\usepackage{amsmath,amssymb}
\usepackage{algorithmic}
\usepackage{graphicx}
\usepackage{xcolor}
\usepackage{tikz}
\usepackage{listings,balance,lstautogobble}
\usepackage{cleveref}
\usepackage{braket}
\usepackage{colortbl, adjustbox}
\usepackage{float}
\usepackage[normalem]{ulem}
\usepackage{etoolbox}

\AtBeginEnvironment{align}{%
  \setlength{\abovedisplayskip}{3.pt}%
  \setlength{\belowdisplayskip}{3.pt}
}

\newcommand{\system}{MERA\xspace}

\Crefname{figure}{Fig.}{Figs.}
\Crefname{equation}{Eq.}{Eq.}
\Crefname{subsection}{Sec.}{Sec.}
\Crefname{section}{Sec.}{Sec.}

\copyrightyear{2026}
\acmYear{2026}
\setcopyright{cc}
\setcctype{by}
\acmConference[ICCAD '26]{IEEE/ACM International Conference on Computer-Aided Design}{November 08--12, 2026}{San Jose, CA, USA}
\acmBooktitle{IEEE/ACM International Conference on Computer-Aided Design (ICCAD '26), November 08--12, 2026, San Jose, CA, USA}
\acmDOI{10.1145/3831252.3833959}
\acmISBN{979-8-4007-2873-0/2026/11}

\begin{document}

\title{\system: A Mid-Circuit-Measurement Error-Aware Compilation Framework for Quantum Circuits}

\author{Ming Zhong}
\email{mzhong25@cse.cuhk.edu.hk}
\orcid{0009-0002-7814-7523}
\affiliation{%
  \institution{The Chinese University of Hong Kong}
  \city{Hong Kong SAR}
  \country{China}
}

\author{Zhemin Zhang}
\orcid{0009-0001-3742-6027}
\email{hisenzhang01@gmail.com}
\affiliation{%
  \institution{The Chinese University of Hong Kong}
  \city{Hong Kong SAR}
  \country{China}
}

\author{Xiangyu Ren}
\orcid{0009-0006-8914-2188}
\email{xiyu.ren@outlook.com}
\affiliation{%
  \institution{The Chinese University of Hong Kong}
  \city{Hong Kong SAR}
  \country{China}
}

\author{Chenghong Zhu}
\orcid{0009-0008-5613-1030}
\email{zhu854@connect.hkust-gz.edu.cn}
\affiliation{%
  \institution{The Hong Kong University of Science and Technology (Guangzhou)}
  \city{Guangzhou}
\state{Guangdong}
\country{China}
}

\author{Siyuan Niu}
\orcid{0000-0003-4683-381X}
\email{siyuan.niu@ucf.edu}
\affiliation{%
  \institution{University of Central Florida}
  \city{Orlando}
\state{Florida}
  \country{USA}
}

\author{Zhiding Liang}
\authornote{Corresponding Author.}
\orcid{0000-0002-7568-0165}
\email{zliang@cse.cuhk.edu.hk}
\affiliation{%
\institution{The Chinese University of Hong Kong; State Key Laboratory of Quantum Information Technologies and Materials (CUHK)}
\city{Hong Kong SAR}
\country{China}
}

\renewcommand{\shortauthors}{M. Zhong, Z. Zhang, X. Ren, C. Zhu, S. Niu, and Z. Liang}

\begin{abstract}
Mid-circuit measurement (MCM) enables qubit reuse and dynamic control in quantum processors, underpinning resource-efficient algorithms and quantum error correction. However, MCM introduces non-negligible errors, including measurement-induced crosstalk, idling-qubit decoherence, and reset infidelity. Additionally, MCM error rates vary across qubits within the same processor, with some exhibiting substantially higher errors. Existing compilers, including Qiskit-compiler and QR-Map, ignore MCM error variations, leading to fidelity loss in circuits with frequent MCM operations.

In this paper, we present \system, an MCM-error-aware quantum compilation framework that integrates MCM error modeling into the layout, routing, and scheduling stages. We first use the lightweight profiling method to capture a per-qubit MCM error distribution that remains stable for over 24 hours. The distribution guides \system in applying MCM-error-aware layout and routing, while \system further incorporates context-aware dynamic decoupling to improve fidelity.
In evaluation, \system delivers 24.94\% -- 80.48\% average fidelity gains over the Qiskit-compiler (optimization level 3) across different groups in selected benchmarks without additional SWAP overhead. It also boosts the fidelity of QR-Map-generated circuits by 29.26\% on average (up to 122.58\%), demonstrating its effectiveness for circuits with frequent MCM operations.
\end{abstract}

\keywords{Quantum Compilation, Middle-Circuit Measurement}

\ccsdesc[500]{Hardware~Quantum computation}

\maketitle

\input{ch1_intro}

\input{ch2_background}
\input{ch3_error_chara}
\input{ch4_compiler_design}
\input{ch5_evaluation}
\input{ch6_related_work}
\input{ch7_conclusion}

\newpage

\bibliographystyle{ACM-Reference-Format}
\bibliography{custom}

\end{document}

%% file: ch1_intro.tex
\section{Introduction}\label{sec:intro}

\begin{figure}[t]
    \centering
    \begin{subfigure}{\linewidth}
        \centering
        \includegraphics[width=\linewidth]{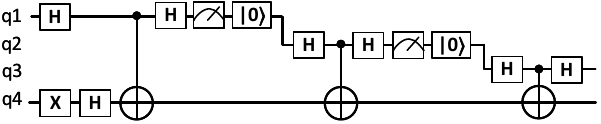}
        \caption{A qubit-reuse circuit for Bernstein-Vazirani (BV) algorithm.}
        \label{fig:bv_circuit}
    \end{subfigure}
    \\
    \begin{subfigure}{\linewidth}
        \centering
        \includegraphics[width=\linewidth]{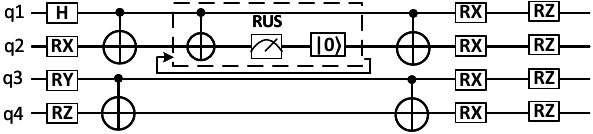}
        \caption{A 4-qubit circuit with repeat-until-success (RUS) operation.}
        \label{fig:rus_circuit}
    \end{subfigure}
    \caption{Representative circuits containing MCM operations.}
    \Description{Representative circuits containing MCM operations.}
    \label{fig:repre_circuits}
\end{figure}

Quantum computing is regarded as a promising paradigm for solving classically intractable problems~\cite{PhysRevLett_qa, quantum_chem,quantum_finance}. However, current quantum hardware remains in the Noisy Intermediate-Scale Quantum (NISQ) era, where limited qubit counts and multiple noise sources hinder reliable quantum computation~\cite{NISQ1, NISQ2}. To alleviate qubit scarcity, modern superconducting processors such as IBM~\cite{Web:IBM} and Google quantum processors~\cite{Web:google_quantum} incorporate mid-circuit measurement (MCM) operations, enabling dynamic circuits where qubits can be measured, reset, and reused during computation. Unlike static circuits, dynamic circuits perform conditional operations based on MCM outcomes, improving resource efficiency and supporting error correction without increasing qubit count. MCM also serves as an important enabler for fault-tolerant quantum error correction, where stabilizer measurements and ancilla resets are repeatedly performed. Moreover, MCM-based conditional control and qubit recycling have been proposed to facilitate distributed quantum computing.
Several studies have leveraged MCM mechanisms, including qubit reset, to enable qubit reuse and improve resource efficiency in quantum computation~\cite{qce_reuse,caqr,qrmap}. A typical example is the Bernstein–Vazirani (BV) circuit~\cite{BV_algo} in \Cref{fig:bv_circuit}, where two MCM stages allow a four-qubit circuit to run on only two qubits, greatly reducing qubit demand. Another representative use is in Repeat-Until-Success (RUS) circuits in \Cref{fig:rus_circuit}, which iteratively execute probabilistic subroutines until a success condition is met, and are widely used in quantum machine learning~\cite{rus1,rus2}.

NISQ superconducting processors exhibit non-negligible MCM errors. Here, an MCM error means that after performing a measurement and then a conditioned reset, the post-reset state is incorrect. These MCM errors arise from several factors, including readout errors, measurement-induced crosstalk, and reset infidelity~\cite{nc_mcm_measure}.
Moreover, unlike neutral-atom qubits, superconducting qubits are artificial in nature, leading to significant device-level variation even within a single chip, such as coherence times, frequency detuning ranges, and other device-level characteristics. Owing to this inherent \textbf{`heterogeneity'} in superconducting processors, we further investigate the qubit-specific variation of MCM errors. As shown in \Cref{fig:eagle_map}, profiling on the 127-qubit IBM Eagle processor shows MCM error rates ranging from 0.05\% to 42.58\% (average 3.42\%). Similarly, results from the 156-qubit IBM Heron processor in \Cref{fig:heron_map} show a range of 0\% to 14.04\% (average 1.19\%).
This variability occurs both across qubits within a device and across different devices; since MCM operations underpin qubit reuse and RUS circuits, mapping MCM qubits to high-error locations can markedly reduce fidelity. Yet mainstream compilers such as Qiskit-compiler~\cite{Web:qiskit_tanspiler} and state-of-art (SOTA) qubit-reuse compilers like CaQR~\cite{caqr} and QR-Map~\cite{qrmap} ignore MCM errors and such outstanding heterogeneity among superconducting qubits, motivating a compilation framework that models MCM error heterogeneity to improve reliability.

In this paper, we first profile MCM errors on real superconducting processors using a simple yet effective method, obtaining a per-qubit error distribution that remains stable for at least 24 hours according to our temporal analysis.
We then propose \system, an MCM-error-aware compiler that incorporates MCM-error modeling into the layout and routing stages, enabling MCM-error-aware qubit mapping and SWAP insertion. During scheduling, it applies as-late-as-possible (ALAP) scheduling and performs context-aware dynamic decoupling (CADD)~\cite{CADD} to mitigate decoherence and measurement-induced crosstalk introduced by MCM operations, thereby further improving fidelity.
While this work focuses on superconducting platforms where qubit `heterogeneity' is most evident,
\system could be extended to other qubit technologies by integrating their hardware-specific characteristics.

To summarize, this work makes the following contributions:

\begin{itemize}[leftmargin=*]
    \item \textbf{An MCM-error-aware compiler} named \system integrates MCM error modeling into layout, routing, and scheduling. We first perform a lightweight profiling to capture a per-qubit MCM error distribution that persists for over 24 hours, avoiding frequent re-profiling. \system leverages this error distribution together with CADD to mitigate MCM induced errors, thereby effectively reducing fidelity loss in circuits involving MCM operations.
    \item \textbf{Comprehensive evaluation} on 29 benchmark circuits, including 8 RUS benchmarks, 19 qubit-reuse benchmarks, and 2 QASMBench benchmarks with MCM operations~\cite{qasmbench}. The Qiskit-compiler (v2.1.2, optimization level 3) serves as the baseline, and QR-Map~\cite{qrmap}, the SOTA qubit-reuse compiler, is included for comparison. Evaluation results show that \system improves circuit fidelity by 24.94\% – 80.48\% on average over the Qiskit-compiler without adding additional SWAP overhead. By post-processing circuits generated by QR-Map, \system further enhances fidelity by 29.26\% on average, confirming \system's effectiveness.
\end{itemize}

%% file: ch2_background.tex
\section{Background}\label{sec:back}

\subsection{Middle-circuit Measurement}\label{sec:mcm_reset}

\begin{figure*}[t]
    \centering
    \begin{subfigure}{.49\linewidth}
        \centering
        \includegraphics[width=\linewidth]{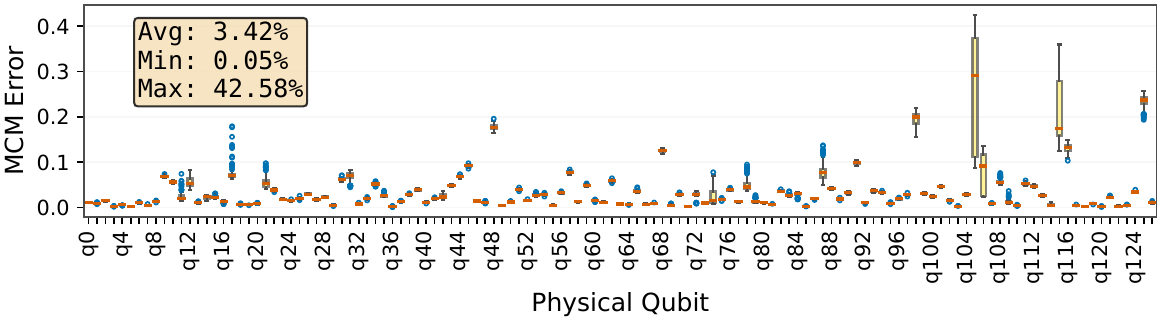}
        \caption{The IBM Eagle quantum processor.}
        \label{fig:eagle_map}
    \end{subfigure}%
    \begin{subfigure}{.49\linewidth}
        \centering
        \includegraphics[width=\linewidth]{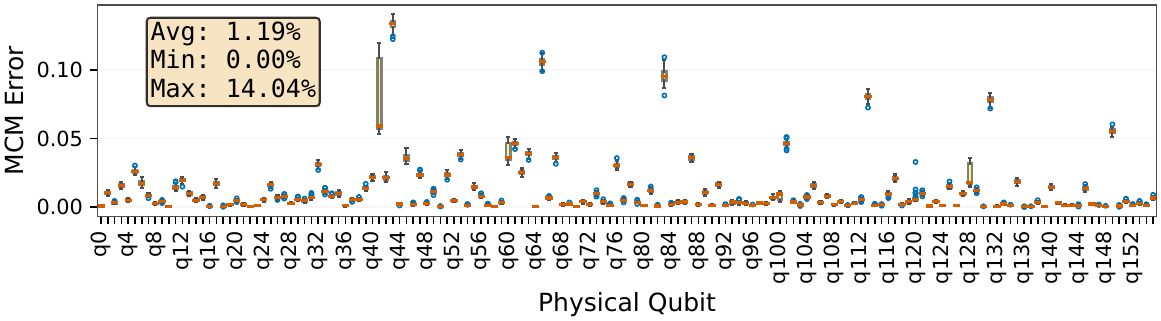}
        \caption{The IBM Heron quantum processor.}
        \label{fig:heron_map}
    \end{subfigure}
    \caption{MCM error distributions for two quantum processors (IBM Eagle and Heron).}
    \Description{MCM error distributions for quantum processors.}
    \label{fig:MCM_map}
    \vspace{-.ex}
\end{figure*}

\textbf{Mid-circuit measurement (MCM)} performs measurements during circuit execution rather than only at the end of the circuit, enabling conditional resets based on intermediate outcomes. By collapsing and re-initializing qubit states, MCM reduces circuit entropy and underpins fault-tolerant quantum computing and dynamic circuits with qubit reuse~\cite{nc_mcm_measure}. After measurement, a qubit can be \textbf{reset} to a known state (typically $\ket{0}$) and reassigned to serve as another logical qubit, thereby improving hardware utilization~\cite{qce_reuse}.

In \Cref{fig:bv_circuit}, a 4-qubit BV circuit is optimized via MCM–based qubit reuse. By measuring and resetting $q_1$ after each two-qubit gate with $q_4$, the same qubit is reused three times, eliminating $q_2$ and $q_3$ and reducing qubit count. Nevertheless, effective qubit reuse relies on high-fidelity MCM operations; otherwise, frequent MCMs can significantly degrade overall circuit fidelity.

\subsection{Error sources of MCM}\label{sec:mcm_errors}

Errors introduced by MCM can be attributed to several sources. First, measurement-induced crosstalk can disturb neighboring qubits. Second, during the long measurement window, unmeasured qubits remain idle and accumulate decoherence errors. Third, the measurement process itself can yield readout errors, misreporting qubit states~\cite{nc_mcm_measure}. For reset operations within MCM, the process involves driving the qubit from $\ket{1}$ to $\ket{0}$ through a bit-flip transition. Reset errors primarily stem from failures in this transition, where the qubit is not successfully restored to the expected ground state.

Our experiments reveal substantial variation in MCM errors across physical qubits on 127- and 156-qubit IBM devices: while some qubits exhibit near-perfect fidelity (MCM error $< 0.1\%$), others show much higher MCM error rates up to 42.58\%, underscoring the need for MCM-error-aware layout and compilation strategies.

Moreover, prior work shows that the fidelity of reset operations strongly depends on the qubit’s pre-reset state: the greater the pre-reset state overlaps with $\ket{1}$, the lower the post-reset fidelity, and vice versa. Applying a second reset pulse typically improves fidelity, while additional repetitions yield diminishing returns, with the optimal number of pulses varying across qubits. Concurrent resets contribute negligibly to total error, indicating that reset infidelity is mainly dominated by qubit-specific imperfections~\cite{qce_reuse}. However, the limitation of this prior work ~\cite{qce_reuse} is that it attributes MCM errors entirely to reset errors, while neglecting the influence of readout, crosstalk, and decoherence errors within MCM.

\subsection{Motivation}\label{sec:mot}

\begin{figure*}[t]
    \centering
    \begin{subfigure}{.49\linewidth}
        \centering
        \includegraphics[width=\linewidth]{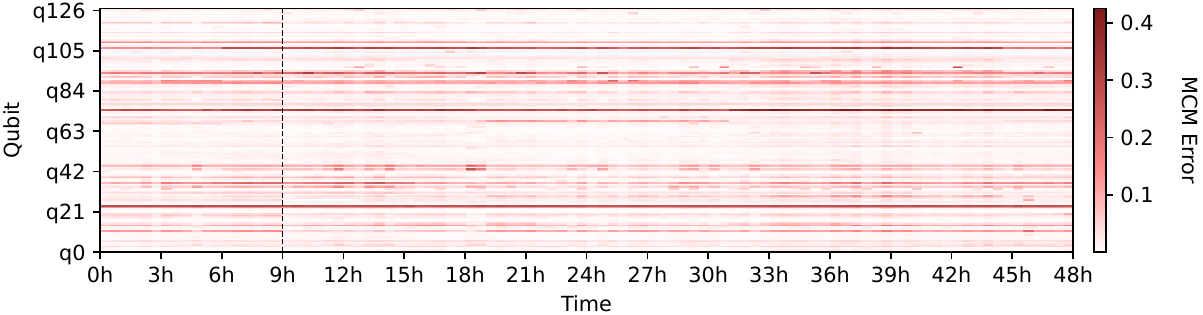}
        \caption{The IBM Eagle quantum processor.}
        \label{fig:egale_temp}
    \end{subfigure}
    \begin{subfigure}{.49\linewidth}
        \centering
        \includegraphics[width=\linewidth]{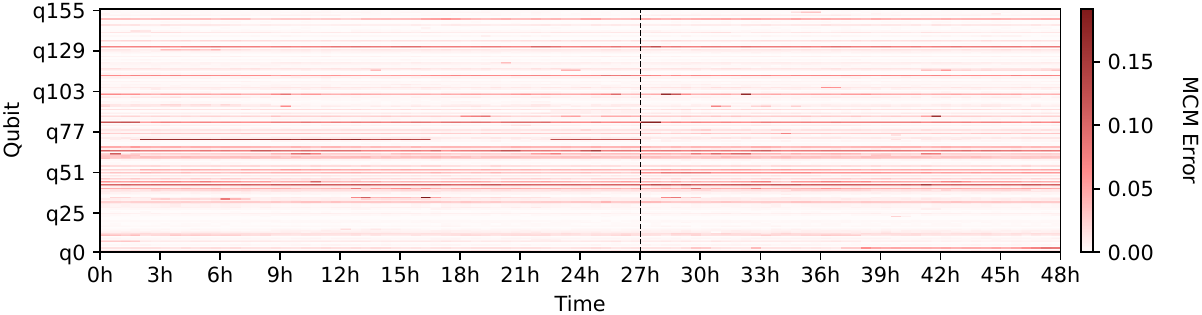}
        \caption{The IBM Heron quantum processor.}
        \label{fig:heron_temp}
    \end{subfigure}
    \caption{Temporal stability of MCM errors in two quantum processors (IBM Eagle and Heron).}
    \Description{Stability of MCM errors in quantum processors.}
    \label{fig:temp_map}
     \vspace{-.ex}
\end{figure*}

MCM is particularly essential in scalable algorithms involving error correction and distributed quantum computing emphasized by qubits reuse. However, MCM introduces errors from MCM-induced crosstalk, decoherence, readout error, and reset infidelity (leakage included). Based on our experiments on real superconducting processors, the IBM devices, we observe significant heterogeneity in MCM errors across qubits due to device-level characteristics.
This observation highlights the need for an MCM-error-aware compiler. Existing works~\cite{qrmap,caqr,qce_reuse} on reset-aware compilation mainly optimize qubit reuse but overlook spatial MCM error variations. Blindly mapping high-intensity MCM operations to high-error qubits can severely degrade fidelity and reliability. Our work address the gap by analyzing MCM error and integrating the awareness and mitigation of different parts of MCM errors into layout, routing, and scheduling to enhance the robustness of quantum compilation.

%% file: ch3_error_chara.tex
\section{MCM Error Profiling}\label{sec:err_chara}

\subsection{Profiling Approach}\label{sec:chara_app}

To evaluate MCM errors on large superconducting processors with minimal execution overhead, we design a simple, device-agnostic profiling approach that applies the following operation sequence to all qubits \emph{in parallel}, thereby directly profiling each qubit’s post-MCM (with reset) success probability $P(\text{Measurement}=|0\rangle)$:
\[
|0\rangle \;\xrightarrow{X\ Gate}\; |1\rangle \;\xrightarrow{MCM\ with\ Reset}\; 
\text{Measurement} \;\Rightarrow\; P(|0\rangle)
\]

Specifically, each qubit is first initialized to the $\ket{1}$ using an X gate, then middle-circuit-measured, reset, and finally measured at the end. The experiment is  repeated 1,024 shots using Qiskit’s Session API to enable large-scale \emph{parallel profiling} and ensure consistent sampling across all qubits. Prior results~\cite{qce_reuse} (discussed in \Cref{sec:mcm_errors}) show that the concurrent (parallel) reset does not induce cross-qubit interference and therefore does not affect our profiling results.

While the existing MCM error profiling method (QIRB~\cite{nc_mcm_measure}) provides precise MCM error rates and can isolate crosstalk-induced errors, its application to large-scale quantum devices incurs substantial experimental cost. In contrast, our profiling method efficiently profiles per-qubit MCM error distributions with far lower cost. Since \system does not require highly precise or frequently updated MCM errors, minor daily fluctuations are tolerable and have negligible impact on compilation quality. Consequently, QIRB is more suitable for scenarios and applications that demand persistently fine-grained error characterization and explicit crosstalk isolation.

\subsection{Profiling Results}\label{sec:res_127}

We perform full-device MCM error profilings on the 127-qubit IBM Eagle and 156-qubit IBM Heron processors to obtain the per-qubit MCM error distribution. \Cref{fig:MCM_map} presents the profiling results, where the orange line denotes the average MCM error of each qubit. On the Eagle processor (\Cref{fig:eagle_map}), MCM errors range from 0.05\% to 42.58\%, with a mean of 3.42\%, revealing substantial spatial variability. Several qubits (e.g., 105 and 115) exhibit notably higher errors, likely due to less frequent calibration or stronger coupling to readout resonators.
A similar pattern appears on the 156-qubit Heron processor (\Cref{fig:heron_map}), where errors span 0\% to 14.04\% with an average of 1.19\%. As a newer generation, Heron benefits from improved flux-line calibration and reduced $ZZ$ crosstalk via tunable couplers, resulting in lower overall MCM errors. Nevertheless, the per-qubit variance remains significant and cannot be ignored.

\subsection{Temporal Stability of MCM Errors}\label{sec:tem_scala}

To evaluate the temporal stability of MCM errors, we repeatedly profiled the IBM Eagle and Heron processors for 48 hours at 0.5-hour intervals and visualized the results as time-series heatmaps (\Cref{fig:temp_map}). Inspection of the profiling log confirmed that the profiling intervals remained essentially 0.5 hours and were not affected by long-running jobs from other users.

As shown in \Cref{fig:egale_temp}, certain qubits (e.g., $q_{74}$, $q_{106}$) consistently exhibit high MCM error rates, while others (e.g., $q_{47}$–$q_{64}$) remain low, demonstrating strong heterogeneity. From hour 9 onward, the overall distribution remains largely stable for over 24 hours, with only minor fluctuations (e.g., at $q_{43}$), and the qubits showing these changes already had elevated errors at hour 9. Thus, a single profiling run provides a reliable MCM error distribution for at least 24 hours, avoiding frequent re-profiling.
A similar trend appears in \Cref{fig:heron_temp}: over the 0–27 hour period, the Heron processor’s MCM error distribution remains stable and is consistently lower than Eagle’s, consistent with the profiling results in \Cref{fig:MCM_map}. These temporal results confirm that MCM errors on superconducting processors remain sufficiently stable for 24 hours or longer, highlighting the value of MCM-error-aware compilation to improve the fidelity of circuits with frequent MCM operations.
Certainly, a shorter profiling interval, such as 4--7 hours, enables more aggressive tracking of MCM drift, but requires more frequent profiling.

\subsection{Device Variation on MCM Errors}\label{sec:device_variation}

The profiling results in the previous sections reveal three important characteristics of MCM errors on current quantum devices. Especially, the superconducting quantum device are artificial qubits, device variation could appear on different hardware parameters like T1, T2, frequency detuning, and many others, in this section, we are majorly investigating on MCM error variation among qubits.

\begin{itemize}[leftmargin=*]
    \item First, MCM error rates vary significantly across devices. As shown in \Cref{fig:MCM_map}, the IBM Eagle processor has MCM errors ranging from 0.05\% to 42.58\% with an average of 3.42\%, while the newer IBM Heron processor reduces this range to 0\% -- 14.04\% with an average of 1.19\%. This difference reflects device-level variations caused by hardware calibration quality, coupling architecture, and improvements in readout and reset mechanisms. And it also highlights the need to profile the target device and incorporate its MCM error distribution into compilation.

    \item Second, even within the same quantum processor, MCM error rates differ significantly across physical qubits. Our profiling results in \Cref{fig:MCM_map} show that some qubits consistently exhibit much higher MCM error rates than others. This spatial heterogeneity implies that blindly mapping MCM-intensive logical qubits to arbitrary physical qubits may lead to severe fidelity degradation. In particular, when circuits contain frequent MCM operations such as qubit reuse or repeat-until-success (RUS) structures, mapping these operations to high-error physical qubits can significantly amplify the overall execution error.

    \item Third, despite the spatial heterogeneity, MCM error distributions exhibit strong temporal stability. As demonstrated in \Cref{fig:temp_map}, the relative error levels of physical qubits remain largely consistent over a period exceeding 24 hours, with only minor fluctuations. This temporal stability suggests that the MCM error distribution obtained from a single profiling run can serve as a reliable reference for compilation during that time window.

\end{itemize}

These observations highlight the need to incorporate qubit-level MCM error information into compilation. Since MCM errors vary across devices and qubits but remain stable over a sufficiently long time window, it is feasible and beneficial to design an MCM-error-aware compiler. Such a compiler can map MCM-intensive logical qubits to lower-error physical qubits while balancing other hardware constraints, thereby improving the reliability of dynamic quantum circuits.

%% file: ch4_compiler_design.tex
\section{\system: An MCM-Error-Aware Compiler}\label{sec:mireq}

\begin{figure*}[t]
\centering
\includegraphics[width=\linewidth]{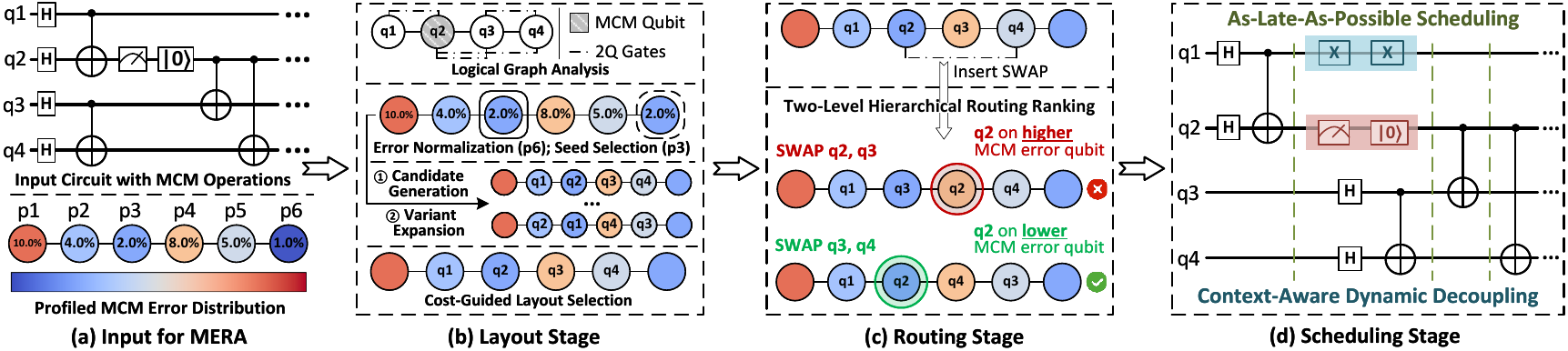}
\caption{The workflow of \system. \emph{More MCMs (on $q_2$), and single- and two-qubit gates (on $q_1$ -- $q_4$) are omitted for brevity.}}
\Description{The workflow of \system.}
\label{fig:workflow}
 \vspace{-.ex}
\end{figure*}

\subsection{Overview of \system}\label{sec:overview}

\Cref{fig:workflow} summarizes the workflow of \system. Given a quantum circuit and the profiled MCM error distribution of the target device (\Cref{fig:workflow}(a)), \system's compilation process goes through three stages: layout, routing, and scheduling.

In the layout stage (\Cref{fig:workflow}(b)), \system defines a cost function that incorporates MCM errors along with distance, two-qubit (2Q), single-qubit (1Q), and readout (RO) errors to guide layout candidate generation and selection.
During routing (\Cref{fig:workflow}(c)), \system employs a two-level ranking strategy that jointly considers routing distance and MCM errors to guide SWAP insertion. This design prioritizes routing decisions that both reduce interaction distance and avoid mapping MCM-intensive logical qubits to high-error physical qubits.
For scheduling (\Cref{fig:workflow}(d)), \system applies ALAP scheduling followed by CADD to improve fidelity of circuits.

\subsection{Layout Stage}\label{sec:layout}

\subsubsection{Logical Graph Analysis}

As shown in \Cref{fig:workflow}(b), \system first analyzes the input circuit and constructs a logical interaction graph $G=(V,E)$, where each logical qubit $q_i \in V$ is a node and each two-qubit gate defines an edge  $(q_i, q_j) \in E$. From the circuit schedule, we also extract the MCM operation intensity (the number of MCM operations) $MCM\_Int(q_i)$ of each qubit $q_i$.
\begin{align}
MCM\_Cost(q_i, p_j) = MCM\_Int(q_i) \times Err_{\text{MCM}}(p_j)
\label{eq:MCM_error}
\end{align}

Additionally, we model the MCM error cost ($MCM\_Cost$) for each logical–physical mapping using \Cref{eq:MCM_error}. When a logical qubit $q_i$ is mapped to a physical qubit $p_j$, its cost is defined as the product of its MCM-operation intensity $MCM\_Int(q_i)$ and the empirical MCM error rate $Err_{\text{MCM}}(p_j)$.
The total circuit-level MCM cost is the sum of these per-qubit MCM costs, linking how frequently each logical qubit performs MCM to the reliability of its mapped physical qubit. This MCM cost design encourages MCM-intensive logical qubits to be placed on physical qubits with lower MCM error.

\subsubsection{MCM Error Normalization and Seed Selection}\label{sec:seed_selection}

In this step, \system selects $N_{Seed}$ physical qubits as seeds for subsequent layout generation.
Instead of selecting layout seeds randomly, \system ranks physical qubits using a normalized composite score that captures error characteristics, including MCM error, two-qubit gate error, single-qubit gate error, readout error, and structural connectivity.
\begin{align}
\widetilde{Err}_{\text{MCM}}(p) =
\begin{cases}
\tau_{\text{MCM}}, & Err_{\text{MCM}}(p) \leq \tau_{\text{MCM}}
\\[1pt]
Err_{\text{MCM}}(p), & Err_{\text{MCM}}(p) > \tau_{\text{MCM}}
\end{cases}
\label{eq:MCM_norm}
\end{align}

Specifically, \system first normalizes the MCM error rates of all physical qubits using the threshold $\tau_{\text{MCM}}$ in \Cref{eq:MCM_norm}. When a qubit’s MCM error rate falls below this threshold, its contribution to fidelity loss remains nearly unchanged after MCM error normalization. Without normalization, the compiler may overemphasize tiny differences in low-error regions and underweight other factors such as distance. MCM error normalization suppresses these sub-threshold variations and restores a balanced weighting among all hardware noise sources in the cost function.

\begin{figure}[t]
\centering
\includegraphics[width=\linewidth]{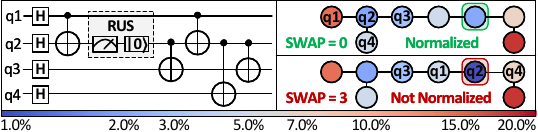}
\caption{Layouts with and without MCM error normalization. The 1.0\% -- 20.0\% range in the error map is \emph{nonlinearly scaled} to highlight subtle differences in color contrast.}
\Description{Layouts with and without MCM error normalization.}
\label{fig:layout_cmp}
 \vspace{-.ex}
\end{figure}

\Cref{fig:layout_cmp} illustrates this effect on a heavy-hex architecture, showing only part of the coupling map for brevity. For the same input circuit, the top layout maps the MCM qubit $q_2$ to a well-connected physical qubit with a 2\% MCM error rate after normalization, while the bottom layout (without normalization) maps it to a less connected qubit with a lower 1\% error rate. Despite the smaller MCM error, the latter requires three additional SWAPs, increasing overhead and reducing fidelity. This highlights the importance of MCM error normalization for balancing reliability and connectivity.
\begin{align}
Score(p) &= \alpha \bigl(1 - \widetilde{Err}_{\text{MCM}}(p)\bigr)
     + \beta \bigl(1 - \bar{e}_{2Q}(p)\bigr)
     + \gamma \bigl(1 - \bar{e}_{1Q}(p)\bigr) \nonumber\\
     &\quad + \delta \bigl(1 - \bar{e}_{RO}(p)\bigr)
     + \epsilon \, {Conn}(p)
\label{eq:seed_score}
\end{align}

After MCM error normalization, \system ranks all physical qubits using \Cref{eq:seed_score} and selects the top $N_{\text{Seed}}$ as layout seeds. $N_{\text{Seed}}$ is determined by the number of qubits in the circuit, thereby dynamically adjusting the number of layout seeds according to the circuit scale.
The scoring function combines connectivity ${Conn}(p)$ (in-degree) with weighted MCM, two-qubit, single-qubit, and readout error rates ($\bar{e}_{2Q}/_{1Q}/_{RO}$) to form a unified score, prioritizing qubits that offer both low error and high connectivity. For example, in \Cref{fig:workflow}(a), qubit $p_6$ with a 1.0\% MCM error is normalized to 2.0\% in \Cref{fig:workflow}(b) and $p_3$ is selected as the seed due to its better connectivity.

\subsubsection{Layout Candidate Generation and Expansion} 

From each seed, \system performs a breadth-first (BFS) expansion over the hardware coupling graph, exploring candidate physical qubits within 1–2 hops (i.e., direct and second-layer neighbors). 
Each candidate placement is evaluated using a layout cost function (\Cref{eq:layout_cost}) that incorporates a distance-aware look-ahead (LA) heuristic together with hardware error metrics. The look-ahead component estimates the interaction distance of upcoming two-qubit gates within several future layers, avoiding short-sighted layout decisions.
\begin{align}
Cost_{\text{Layout}}(L_M) = w_{\text{dist}}\,\mathrm{LA}(L_M, look\_ahead) +  \sum_{(q_i, p_j)\in L_{M}} \nonumber \\[-3pt]
 \Big[
    w_{\text{MCM}}\, MCM\_Cost(q_i, p_j)
  + w_{2Q}\, N^{\text{2Q}}_{p_j}(L_M)\,\bar{e}_{2Q}(p_j) + \nonumber \\[-3pt]
   w_{1Q}\, N^{\text{1Q}}_{p_j}(L_M)\,\bar{e}_{1Q}(p_j)
  + w_{RO}\, N^{\text{RO}}_{p_j}(L_M)\,\bar{e}_{RO}(p_j)
\Big]
\label{eq:layout_cost}
\end{align}

In \Cref{eq:layout_cost}, $L_M$ denotes the current logical-to-physical qubit mapping, and the $LA$ function represents the look-ahead distance heuristic, where $look\_ahead$ specifies the number of future layers considered. We extend this cost function by incorporating MCM errors and the 2Q/1Q/readout error rates ($\bar{e}{2Q}$/$\bar{e}{1Q}$/$\bar{e}{RO}$) of each logical–physical pair $(q_i, p_j)$ in $L_M$, weighted to form a composite cost function. Here, $N^{\text{2Q/1Q/RO}}{p_j}(L_M)$ denotes the number of corresponding 2Q/1Q/readout operations executed on physical qubit $p_j$ under the layout mapping $L_M$.

During the BFS expansion for layout candidate generation, let $I$ denote the current partial mapping and $(q_u, p_v)$ the next logical–physical qubit pair to be mapped. \Cref{eq:delta_cost} then evaluates the incremental cost of extending the current mapping $I$ by $(q_u, p_v)$  by applying \Cref{eq:layout_cost}, guiding the BFS expansion for layout generation.
\begin{align}
\Delta C((q_u, p_v) \mid I) &= Cost_{\text{Layout}}(I \cup (q_u, p_v)) - Cost_{\text{Layout}}(I)
\label{eq:delta_cost}
\end{align}

The BFS expansion terminates once all logical qubits are mapped to distinct physical qubits, or when no lower-cost neighbor within the 2-hop radius can further reduce the incremental cost $\Delta C$. If termination occurs before full coverage, \system optionally performs a single shortest-path bridging step to connect the remaining unmapped qubits to the partial layout, ensuring a fully connected and valid mapping.
After generating initial candidate mappings, \system increases layout diversity by generating intra-candidate variants, i.e., reassigning logical–physical pairs within the same selected physical set. For each candidate, variants are created using two reordering rules:
(1) MCM-error-aware reordering: assign logical qubits with higher MCM intensity to physical qubits with lower MCM error;
(2) Connectivity-aware reordering: assign logical qubits with higher two-qubit participation to physical qubits with higher connectivity centrality in the hardware coupling graph.

This variant generation broadens the local search around each BFS expansion, improving robustness against local minima and increasing the likelihood of finding a layout with lower cost.

\subsubsection{Cost-Guided Layout Selection}

For all generated layout candidates, \system computes the global layout cost again using \Cref{eq:layout_cost} and selects the layout with the lowest cost as the final layout.

\subsection{Routing Stage}\label{sec:routing}
In the routing stage, \system adopts a distance-guided SWAP insertion strategy enhanced with a two-level hierarchical ranking mechanism to balance routing efficiency and MCM reliability.

The routing algorithm generates candidate SWAP operations from coupling edges adjacent to the currently interacting qubits, preserving locality in the hardware topology.
When no candidate yields a shorter logical-to-physical distance, the routing algorithm falls back to the previous candidate set to maintain routing progress and prevent stalling.

Next, \system employs a two-level hierarchical ranking strategy to evaluate and prioritize SWAP candidates, as shown in \Cref{fig:workflow}(c). 

\begin{itemize}[leftmargin=*]
\item \textbf{Level-1 Distance Cost}. The primary ranking metric evaluates the reduction in interaction distance for both current and near-future 2Q gates within a limited look-ahead window. Specifically, \system computes a weighted distance cost based on the current 2Q interaction and a ${look\_ahead}$ window of upcoming 2Q gates. The accumulated distance serves as the primary ranking metric, favoring candidates that reduce near-term routing overhead.

\item \textbf{Level-2 MCM Error}. For candidates whose Level-1 scores lie within the $\Delta_{\text{swap}}$ range, \system further ranks SWAPs by the product of the post-swap physical qubit’s MCM error rate and the logical qubit’s remaining MCM intensity after routing. The lowest-product candidate is selected, preventing high-intensity logical qubits from mapping to high-error physical qubits.
\end{itemize}

As shown in \Cref{fig:workflow}(c), qubit $q_2$ with more MCM operations must interact with $q_4$ via a 2Q gate. A SWAP can be inserted between $q_2$ -- $q_3$ or  $q_3$ -- $q_4$. The latter is selected because it prevents the MCM-intensive qubit $q_2$ from being relocated to a physical qubit with higher MCM error rates. Additional MCM operations on $q_2$ are omitted for brevity but explained in the caption of \Cref{fig:workflow}.

\subsection{Scheduling Stage}\label{sec:scheduling}

After routing, \system applies an as-late-as-possible (ALAP) scheduler to compact gate execution and explicitly expose idle intervals. A Context-Aware Dynamic Decoupling (CADD) module then inserts decoupling pulses only within these ALAP-generated delay windows to suppress decoherence.

Unlike conventional DD, CADD adapts its pulse placement to circuit context, including MCM timing and qubit-dependent noise, thereby avoiding MCM windows and minimizing interference with subsequent gates. By coordinating pulse timing across coupled qubits, CADD also mitigates the $ZZ$ crosstalk that dominates during MCM operations~\cite{nc_mcm_measure}. As illustrated in \Cref{fig:workflow}(d), ALAP creates a delay window on $q_1$ during the MCM on $q_2$, while $q_3$ and $q_4$ are shifted rightward so that their final 1Q and 2Q gates align with the end of the MCM window. Consequently, CADD inserts $X$–$X$ sequences only on $q_1$, with no need for DD on $q_3$ or $q_4$.
By combining ALAP scheduling with CADD, \system provides fine-grained noise suppression tailored to MCM-induced idle behavior.

%% file: ch5_evaluation.tex
\section{Evaluation}\label{sec:eva}

\subsection{Evaluation Setup}\label{sec:eva_setpup}
\input{Table/parameters}

\begin{figure*}[t]
    \centering
    \begin{subfigure}{.49\linewidth}
        \centering
        \includegraphics[width=\linewidth]{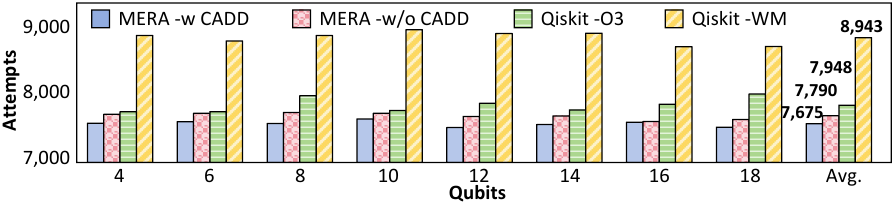}
        \caption{Comparison of the \textbf{Attempts} metric. \textbf{Lower} is \textbf{better}.}
        \label{fig:rus_attempts}
    \end{subfigure}
    \begin{subfigure}{.49\linewidth}
        \centering
        \includegraphics[width=\linewidth]{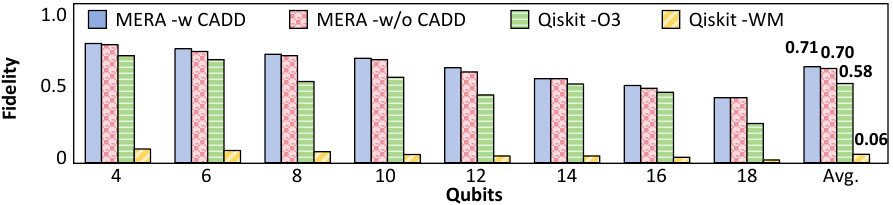}
        \caption{Comparison of the \textbf{Avg. Fidelity} metric. \textbf{Higher} is \textbf{better}.}
        \label{fig:rus_fidelity}
    \end{subfigure}
    \caption{Evaluation results of eight RUS benchmarks. Note that the average critical depth (\textbf{Path} = 28) and average number of inserted SWAP gates (\textbf{SWAP} = 0) are \textbf{identical} for \system and Qiskit under different options, and thus omitted for brevity.}
    \Description{Evaluation results of eight RUS benchmarks.}
    \label{fig:rus}
    \vspace{-.ex}
\end{figure*}

\textbf{Implementations.} We implemented \system on Qiskit v2.1.2. The parameter settings described in \Cref{sec:mireq} are listed in \Cref{table:parameters}. These parameter values were determined through grid-search tuning over the benchmarks in \Cref{fig:rus} and \Cref{table:reuse}, balancing compilation overhead and fidelity improvement on selected benchmarks.

\begin{figure}[t]
    \centering
    \begin{subfigure}{\linewidth}
        \centering
        \includegraphics[width=\linewidth]{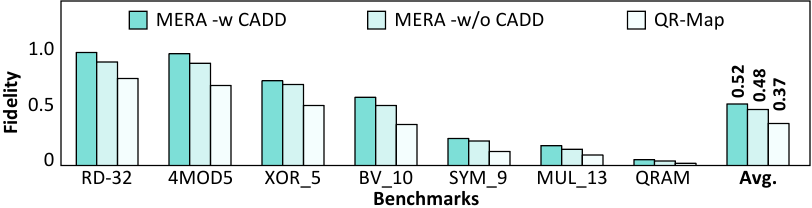}
        \caption{The IBM Eagle quantum processor.}
        \label{fig:egale_exp}
    \end{subfigure}
    \\
    \begin{subfigure}{\linewidth}
        \centering
        \includegraphics[width=\linewidth]{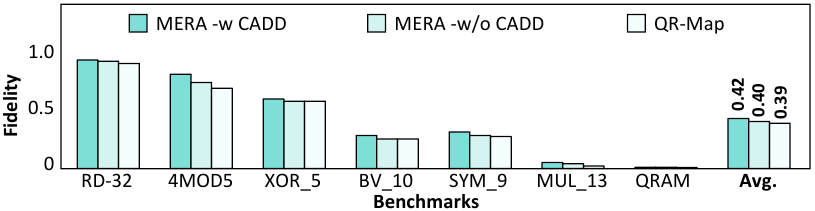}
        \caption{The IBM Heron quantum processor.}
        \label{fig:heron_exp}
    \end{subfigure}
    \caption{Evaluation of benchmarks on real devices.}
    \Description{Fidelity evaluation of benchmark circuits on real devices.}
    \label{fig:real_machine}
    \vspace{-.ex}
\end{figure}

\begin{figure*}[t]
    \centering
    \begin{subfigure}{.33\linewidth}
        \centering
        \includegraphics[width=\linewidth]{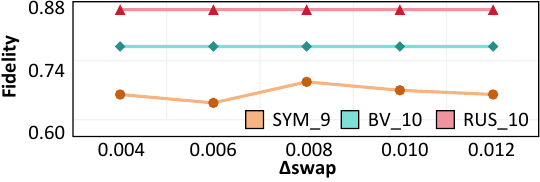}
        \caption{$\Delta_{swap}$}
        \label{fig:swap}
    \end{subfigure}
    \begin{subfigure}{.33\linewidth}
        \centering
        \includegraphics[width=\linewidth]{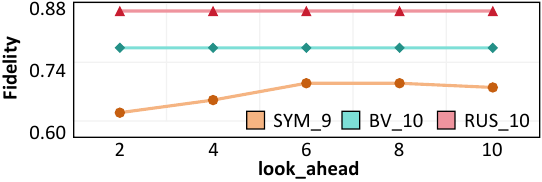}
        \caption{$look\_{ahead}$}
        \label{fig:lookahead}
    \end{subfigure}
    \begin{subfigure}{.33\linewidth}
        \centering
        \includegraphics[width=\linewidth]{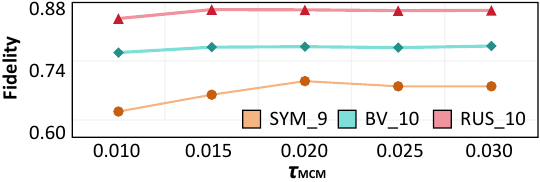}
        \caption{$\tau_{MCM}$}
        \label{fig:tau}
    \end{subfigure}
    \caption{Sensitivity analysis of three parameters ($\Delta_{swap}$, $look_{ahead}$, and $\tau_{MCM}$) listed in \Cref{table:parameters}.}
    \Description{Evaluation results of eight RUS benchmarks.}
    \label{fig:sensi}
    \vspace{-.ex}
\end{figure*}

\begin{figure}[t]
    \centering
    \includegraphics[width=\linewidth]{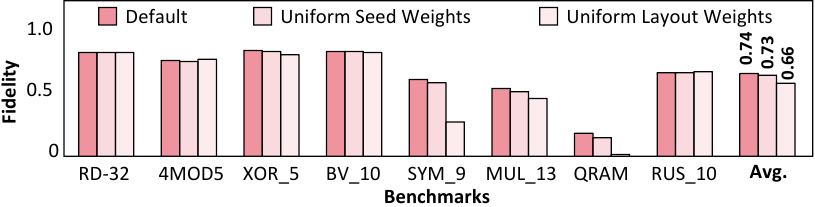}
    \caption{Ablation study of the weights setting in \Cref{table:parameters}.}
    \label{fig:abal}
    \Description{Ablation study of the weights setting}
    \vspace{-.ex}
\end{figure}

\noindent
\textbf{Baselines.}
We use the Qiskit-compiler (v2.1.2) as a baseline, which does not model MCM errors. Two Qiskit settings are evaluated: \textbf{Qiskit -O3} (optimization level 3) and \textbf{Qiskit -WM} (worst MCM error mapping), which constrains the layout to a contiguous physical subgraph with the highest cumulative MCM cost (MCM-intensity × MCM-error in \Cref{eq:MCM_error}) from the measured MCM distribution. We also include QR-Map~\cite{qrmap}, a SOTA qubit-reuse compiler, to assess fidelity gains when using \system to post-process its outputs.

\noindent
\textbf{Evaluation Platforms.}
We conduct the evaluations in \Cref{sec:cmp_qiskit,sec:cmp_qrmap} using a simulator that replicates the 127-qubit IBM Eagle coupling map and incorporates all noises, including MCM error rates, extracted from the calibration data in \Cref{sec:err_chara}. The simulator is implemented with Qiskit’s AerSimulator in "statevector" mode to support dynamic circuits, with $CX$ as the native two-qubit gate. For real-device evaluation (\Cref{sec:cmp_real}), we use the 127-qubit IBM Eagle and 156-qubit IBM Heron processors.

\noindent
\textbf{Benchmarks.}
We select 29 benchmarks for evaluation, including 8 RUS benchmarks with qubit counts from 4 to 18 in steps of 2. A 4-qubit RUS benchmark used in our evaluation is shown in \Cref{fig:rus_circuit}. All 8 RUS benchmarks are designed to place 2Q gates on adjacent qubits (e.g., $q_1$–$q_2$, $q_3$–$q_4$, ...) before and after the RUS segment, ensuring no SWAPs are needed under proper layouts, therefore making MCM-induced fidelity loss directly observable by isolating SWAPs.
Additionally, our evaluation includes 8 qubit-reuse circuits from existing work~\cite{qce_reuse}, 2 benchmarks with MCMs from QASMBench~\cite{qasmbench}, and 11 qubit-reuse circuits generated by QR-Map~\cite{qrmap}.

\noindent
\textbf{Evaluation Metrics.}
For each benchmark, we run 1,024 shots per iteration and repeat each benchmark for five iterations. For RUS circuits, we define the \textbf{Attempts} metric as the total number of MCM operations required for the circuit to reach success across all 5 × 1,024 shots (\textbf{for RUS circuits only}); lower values of Attempts indicate higher MCM success rates. For all benchmarks, the average execution fidelity (\textbf{Fidelity}) is measured using Hellinger fidelity~\cite{fidelity}. We also adopt two metrics from prior work~\cite{qce_reuse, qrmap, caqr} -- the average number of inserted SWAPs (\textbf{SWAP}) and the average critical path length (\textbf{Path}) -- to quantify execution overhead.

\subsection{Comparison with Qiskit-Compiler}\label{sec:cmp_qiskit}

\textbf{RUS Benchmarks.}
As shown in \Cref{fig:rus_fidelity}, \system with CADD consistently outperforms Qiskit -O3 and Qiskit -WM across all eight RUS benchmarks, achieving fidelity improvements of 6.90\%–65.52\% (average 24.94\%) over Qiskit -O3, while requiring fewer attempts (\Cref{fig:rus_attempts}) and introducing neither extra SWAPs nor longer critical paths. Even without CADD, \system still delivers an average fidelity gain of 22.81\%, underscoring the effectiveness of MCM-aware layout and routing. Under Qiskit’s worst MCM error layout (-WM), attempts increase sharply and fidelity collapses, highlighting the strong impact of MCM errors on RUS circuits.

\input{Table/reuse_table}

\noindent
\textbf{Qubit-Reuse and QASMBench Benchmarks.} 
\Cref{table:reuse} presents results on eight qubit-reuse circuits and two QASMBench benchmarks. \system, with and without CADD, achieves average fidelity improvements of 80.48\% and 75.13\% over Qiskit -O3, respectively, and consistently surpasses Qiskit -WM across all cases without adding SWAPs. In the Shor5 benchmark, where only two MCM operations occur, \system shows no improvement over Qiskit -O3, indicating that its benefits are more pronounced for circuits with higher MCM intensity, such as RUS and qubit-reuse circuits.

\input{Table/qrmap}

\subsection{Improvements over QR-Map}\label{sec:cmp_qrmap}

We use \system to post-process 11 qubit-reuse circuits originally generated by QR-Map. As shown in \Cref{table:qrmap}, for the SYM\_9 benchmark, \system achieves fidelity improvements of 122.58\% (with CADD) and 103.23\% (without CADD). \emph{Two larger benchmarks (QAOA 20–10 and QRAM 20–13) are included to assess scalability}, though their fidelities remain low due to their larger qubit counts. Overall, \system achieves average improvements of 29.26\% with CADD and 23.42\% without CADD over QR-Map.
These results underscore the effectiveness of MCM-error-aware layout and routing in improving the reliability of qubit-reuse circuits. Importantly, \system is orthogonal to existing compilation frameworks, including distributed, quantum error correction, and qubit-reuse compilers, and can serve as an add-on module to further improve fidelity in any compiler workflow involving MCM operations. Notably, the compilation runtime of \system ranges from 0.12 to 0.77 seconds on 11 benchmarks. Although this introduces modest runtime overhead, the resulting fidelity improvements make the trade-off worthwhile.

\subsection{Real-Device Evaluation}\label{sec:cmp_real}

To evaluate the scalability of \system, we conducted real-device experiments on the 127-qubit IBM Eagle and 156-qubit IBM Heron processors using 7 benchmark circuits adopted in the real-device evaluation in QR-Map~\cite{qrmap}. As shown in \Cref{fig:egale_exp}, \system achieves an average fidelity improvement of 40.55\% over QR-Map with CADD and 39.73\% without CADD across all benchmarks on Eagle. This fidelity gain demonstrates the scalability and effectiveness of \system.
On the Heron processor (\Cref{fig:heron_exp}), \system also outperforms QR-Map on all benchmarks, but the gains are noticeably smaller (averaging 7.69\% with CADD and 2.56\% without CADD). The smaller gain arises from Heron’s substantially lower MCM error rates compared to Eagle (as shown in \Cref{fig:MCM_map}), leaving limited room for MCM-aware layout and routing to further improve fidelity.

\subsection{Sensitivity Analysis}\label{sec:sensi}

We conduct a sensitivity analysis on three parameters listed in \Cref{table:parameters}, namely $\Delta_{swap}$, $look_{ahead}$, and $\tau_{MCM}$. The other two groups of parameters are evaluated with ablation studies in \Cref{sec:abalation}.

To evaluate parameter robustness, we select three circuits: RUS\_10, BV\_10, and SYM\_9, covering RUS, qubit reuse without SWAP insertion, and qubit reuse with SWAP insertion, respectively. For each parameter ($\Delta_{swap}$, $look_{ahead}$, and $\tau_{MCM}$), we test five values centered around the default setting in \Cref{table:parameters}, including two smaller and two larger values, and analyze the resulting fidelity variations.

As shown in \Cref{fig:sensi}(a), both RUS\_10 and BV\_10 do not require SWAP insertion during the routing stage. Therefore, their fidelities remain unchanged as $\Delta_{swap}$ varies. In contrast, for the SYM\_9 circuit, the fidelity varies only within the range of 0.64 -- 0.69 without significant fluctuations. This result indicates that \system exhibits low sensitivity to the value of $\Delta_{swap}$ within the tested range.

For $look_{ahead}$, \Cref{fig:sensi}(b) shows that the fidelities of RUS\_10 and BV\_10 remain unchanged, as this parameter mainly affects future-interaction prediction during SWAP insertion. For SYM\_9, fidelity improves when $look_{ahead}$ increases from 2 to 6, then stabilizes, with only a slight drop of about 0.01 at 10. Since larger values introduce higher compilation overhead, we set $look_{ahead}=6$ to balance performance and time cost.

Finally, \Cref{fig:sensi}(c) shows the impact of $\tau_{MCM}$. Increasing $\tau_{MCM}$ from 0.01 to 0.015 noticeably improves fidelity across all three circuits, suggesting that overly emphasizing MCM-error differences may cause layout and routing to neglect other fidelity-related factors, such as two-qubit gate errors and qubit distances. When $\tau_{MCM}$ further increases from 0.02 to 0.03, fidelities remain relatively stable, indicating that \system is not sensitive to $\tau_{MCM}$ within this range.

\subsection{Ablation Study}\label{sec:abalation}

We further conduct an ablation study on the two groups of weights listed in \Cref{table:parameters}. In \Cref{eq:seed_score}, the weights of MCM error and two-qubit (2Q) error are set to 0.25, which is slightly higher than the uniform value of 0.2. In \Cref{eq:layout_cost}, the weight of the distance term is set to 0.45, which is significantly larger than the uniform value. To validate the rationale of this design, we construct the uniform weight configuration [0.2, 0.2, 0.2, 0.2, 0.2] as comparison baselines.

Specifically, we denote the parameter configuration in \Cref{table:parameters} as "Default" setting. When only the weights in \Cref{eq:seed_score} are replaced with [0.2, 0.2, 0.2, 0.2, 0.2], the configuration is labeled "Uniform Seed Weights". Similarly, when the weights in \Cref{eq:layout_cost} are replaced with [0.2, 0.2, 0.2, 0.2, 0.2], the configuration is labeled "Uniform Layout Weights". We evaluate these configurations using the seven qubit-reuse benchmarks and one RUS\_10 benchmark from \Cref{sec:cmp_real}.

As shown in \Cref{fig:abal}, the difference between Default Setting and Uniform Seed Weights is relatively small. This is because the Default Setting only assigns slightly higher weights to MCM error and 2Q error compared to the uniform configuration. In contrast, the difference between Default Setting and Uniform Layout Weights is more significant for test cases that require SWAP insertion, especially those with a large number of SWAP operations such as SYM\_9 and MUL\_13. Under the Uniform Layout Weights configuration, the distance term receives much less emphasis, which leads to more SWAP insertions and consequently reduces circuit fidelity.

On average, the Default Setting improves fidelity of selected benchmarks by approximately 0.01 compared with Uniform Seed Weights, and by 0.08 compared with Uniform Layout Weights.

\subsection{Discussion}\label{sec:exp_dis}

Evaluation shows that \system effectively improves benchmark fidelity by mitigating MCM-induced crosstalk, decoherence, readout error, and qubit-dependent reset infidelity. However, our current MCM-error model considers only $\ket{1}\rightarrow\ket{0}$ reset transitions; practical resets also suffer from leakage to higher energy levels, which should be addressed through hardware mechanisms rather than the compiler~\cite{nc_leakage}. Likewise, \system treats readout error only as a cost term, whereas additional mitigation can be achieved via hardware- or protocol-level methods~\cite{dyrem,sprem,koh2025readouterrormitigationmidcircuit}. Future work should integrate hardware leakage suppression, \system, and dedicated readout-error mitigation to further reduce overall MCM-induced errors.

%% file: Table/parameters.tex
\begin{table}[t]
\caption{Parameter settings for the implementation of \system.}
\label{table:parameters}
\setlength{\tabcolsep}{.1ex}
\setstretch{1.}
{
\begin{adjustbox}{max width=\linewidth}
\begin{tabular}{|c|l|c|}
\hline
\multicolumn{1}{|c|}{\textbf{Notation}}                      & \multicolumn{1}{c|}{\textbf{Description}}                                  & \multicolumn{1}{c|}{\textbf{Values}} \\ \hline
$\tau_{MCM}$                                              & Threshold of MCM error rate in \Cref{eq:MCM_norm}.                                             & 2\%                                \\ \hline
{[}$\alpha$, $\beta$, $\gamma$, $\delta$, $\epsilon${]}     & Weights for scoring the seed qubit in \Cref{eq:seed_score}.      & {[}0.25,0.25,0.1,0.2,0.2{]}   \\ \hline
$look\_ahead$                                               & Number of look ahead layers in \Cref{eq:layout_cost}.   & 6                                   \\ \hline
$w_{dist/MCM/2Q/1Q/RO}$ & Weights for the cost function in \Cref{eq:layout_cost}. & {[}0.45,0.2,0.2,0.05,0.1{]} \\ \hline
$\Delta_{swap}$                                             & Threshold for inserting swaps used in \Cref{sec:routing}.                                  & 0.008                               \\ \hline
\end{tabular}
\end{adjustbox}
}
\vspace{-.ex}
\end{table}

%% file: Table/reuse_table.tex
\begin{table}[t]
\caption{Evaluation results of qubit-reuse and QASMBench benchmarks. In the "Qubits" column, $m - n$ denotes circuits with $m$ logical qubits mapped to $n$ physical qubits after reuse.}
\label{table:reuse}
\setlength{\tabcolsep}{.2ex}
\setstretch{1.}
{
\begin{adjustbox}{width=\linewidth}
\begin{tabular}{c|c|ccc|ccc|ccc|ccc}
\hline
\multirow{2}{*}{\textbf{Bench}} & \multirow{2}{*}{\textbf{Qubits}} & \multicolumn{3}{c|}{\textbf{\system \ -w CADD}}           & \multicolumn{3}{c|}{\textbf{\system \ -w/o CADD}}         & \multicolumn{3}{c|}{\textbf{Qiskit -O3}}                             & \multicolumn{3}{c}{\textbf{Qiskit -WM}}                                \\ \cline{3-14} 
                                &                                  & Path & SWAP & Fidelity & Path & SWAP & Fidelity & Path & SWAP & Fidelity & Path & SWAP & Fidelity \\ \hline
\multirow{3}{*}{\shortstack{BV}}             & 4-2                              & 20                                                   & 0    & 0.95     & 20                                                   & 0    & 0.93     & 20                                                   & 0    & 0.69     & 20                                                   & 0    & 0.34     \\
                                & 7-2                              & 38                                                   & 0    & 0.89     & 38                                                   & 0    & 0.88     & 38                                                   & 0    & 0.53     & 38                                                   & 0    & 0.13     \\
                                & 10-2                             & 63                                                   & 0    & 0.86     & 63                                                   & 0    & 0.83     & 63                                                   & 0    & 0.42     & 63                                                   & 0    & 0.04     \\
                                & 20-2                             & 100                                                   & 0    & 0.68     & 100                                                   & 0    & 0.66     & 100                                                   & 0    & 0.26     & 100                                                   & 0    & 0.001     \\
                                & 50-2                             & 250                                                   & 0    & 0.36     & 250                                                   & 0    & 0.33     & 250                                                   & 0    & 0.11     & 250                                                   & 0    & 0.00     \\                               
                                \hline
\multirow{3}{*}{\shortstack{H--\\Ladder}}          & 3-2                              & 24                                                   & 0    & 0.96     & 24                                                   & 0    & 0.95     & 24                                                   & 0    & 0.75     & 24                                                   & 0    & 0.45     \\
                                & 5-2                              & 40                                                   & 0    & 0.93     & 40                                                   & 0    & 0.91     & 40                                                   & 0    & 0.61     & 40                                                   & 0    & 0.22     \\
                                & 7-2                              & 56                                                   & 0    & 0.88     & 56                                                   & 0    & 0.87     & 56                                                   & 0    & 0.50     & 56                                                   & 0    & 0.10     \\ \hline
IPEA                            & 2                                & 37                                                   & 0    & 0.91     & 37                                                   & 0    & 0.91     & 37                                                   & 0    & 0.61     & 37                                                   & 0    & 0.26     \\ \hline
Shor                           & 5                                & 89                                                   & 7    & 0.98     & 89                                                   & 7    & 0.98     & 89                                                   & 7    & 0.98     & 89                                                   & 7    & 0.90     \\ \hline
\end{tabular}
\end{adjustbox}
}
\vspace{-.ex}
\end{table}

%% file: Table/qrmap.tex
\begin{table}[t]
\caption{Fidelity comparison with QR-Map.
}
\label{table:qrmap}
\setlength{\tabcolsep}{.4ex}
\setstretch{1.}
{
\begin{adjustbox}{max width=\linewidth}
\begin{tabular}{c|c|ccc|ccc|ccc}
\hline
\multicolumn{1}{c|}{\multirow{2}{*}{\textbf{Bench}}} & \multirow{2}{*}{\textbf{Qubits}} & \multicolumn{3}{c|}{\textbf{\system \ -w CADD}}            & \multicolumn{3}{c|}{\textbf{\system \ -w/o CADD}}          & \multicolumn{3}{c}{\textbf{QR-Map}}                                    \\ \cline{3-11} 
\multicolumn{1}{c|}{}                                &                                 &  Path &  SWAP &  Fidelity & Path &  SWAP & Fidelity & Path & SWAP & Fidelity \\ \hline
\multirow{2}{*}{QAOA}                                & 10-6                            & 79                                                   & 18   & 0.56     & 79                                                   & 18   & 0.55     & 79                                                   & 18   & 0.53     \\
& 20-10                          & 145                                                  & 61   & 8e-5     & 145                                                  & 61   & 8e-5     & 145                                                  & 61   & 5e-5     \\ \hline
QAOA-L2                           & 10-7                            & 139                                                  & 25   & 0.57     & 139                                                  & 25   & 0.52     & 139                                                  & 25   & 0.50    \\ \hline
QAOA-L3                             & 10-9                            & 167                                                  & 49   & 0.51     & 167                                                  & 49   & 0.49     & 167                                                  & 49   & 0.48     \\ \hline 
RD-32                                                 & 4-3                             & 41                                                   & 3    & 0.95     & 41                                                   & 3    & 0.93     & 41                                                   & 3    & 0.84     \\
4MOD5                                                & 5-3                             & 31                                                   & 4    & 0.94     & 31                                                   & 4    & 0.91     & 31                                                   & 4    & 0.86     \\
XOR\_5                                                 & 6-2                             & 10                                                   & 0    & 0.93     & 10                                                   & 0    & 0.92     & 10                                                   & 0    & 0.75     \\
BV\_10                                                 & 10-2                            & 63                                                   & 0    & 0.86     & 63                                                   & 0    & 0.84     & 63                                                   & 0    & 0.67     \\
SYM\_9                                                 & 12-5                            & 392                                                  & 72   & 0.69     & 392                                                  & 72   & 0.63     & 392                                                  & 72   & 0.31     \\
MUL\_13                                                & 13-7                            & 129                                                  & 32   & 0.61     & 129                                                  & 32   & 0.60     & 129                                                  & 32   & 0.50     \\
QRAM                                                 & 20-13                           & 395                                                  & 180  & 0.21     & 395                                                  & 180  & 0.18     & 395                                                  & 180  & 0.18     \\ \hline
\end{tabular}
\end{adjustbox}
}
\vspace{-.ex}
\end{table}

%% file: ch6_related_work.tex
\section{Related Work}\label{sec:rel_work}

\textbf{Quantum Compilers.}
Quantum compilers~\cite{SQUARE,Paulihedral,ScaffCC,caqr,Web:qiskit_tanspiler,qrmap,fermihedral,TILT} translate abstract algorithms into quantum circuits through techniques such as gate fusion, re-synthesis, and error mitigation. 
Existing compilers do not model MCM errors; \system\ fills this gap by explicitly incorporating MCM errors into the compilation process.

\noindent
\textbf{Quantum Error Mitigation.}
A broad range of methods address NISQ noise and perform quantum error correction~\cite{9773181, dyrem,sprem,readout_error,DAC_error_correction,chen_error}.
Specifically, several studies characterize MCM errors~\cite{nc_mcm_measure,mcm_random,qce_reuse,koh2025readouterrormitigationmidcircuit}, but none incorporate them into the compilation process. \system fills this gap by performing MCM-error-aware compilation.

%% file: ch7_conclusion.tex
\section{Conclusion}\label{sec:conclusion}

This paper presents \system, a compilation framework that incorporates MCM error awareness into layout, routing, and scheduling. By leveraging profiled per-qubit MCM error distributions, \system effectively mitigates MCM-induced fidelity loss, achieving 24.94\% -- 80.48\% average fidelity improvements over the Qiskit-compiler (-O3 setting) and 29.26\% gains over QR-Map across 29 benchmarks, demonstrating  its effectiveness for MCM circuits.

\section*{Acknowledgment}
We thank all anonymous reviewers for their constructive comments. We also thank Hyungseok Kim and Enhyeok Jang for providing the QR-Map-generated circuits used in our evaluation.

%% file: custom.bib
@String{Computing = "Computing" }

@String{Computer = "{IEEE} Computer" }

@String{Springer = "Springer-Verlag" }

@article{PhysRevLett_qa,
  title = {{Quantum Algorithm for Linear Systems of Equations}},
  author = {Harrow, Aram W. and Hassidim, Avinatan and Lloyd, Seth},
  journal = {Phys. Rev. Lett.},
  volume = {103},
  issue = {15},
  pages = {150502},
  numpages = {4},
  year = {2009},
  month = {Oct},
  publisher = {American Physical Society},
  doi = {10.1103/PhysRevLett.103.150502},
  url = {https://link.aps.org/doi/10.1103/PhysRevLett.103.150502}
}

@article{quantum_chem,
author = {Cao, Yudong and Romero, Jonathan and Olson, Jonathan P. and Degroote, Matthias and Johnson, Peter D. and Kieferová, Mária and Kivlichan, Ian D. and Menke, Tim and Peropadre, Borja and Sawaya, Nicolas P. D. and Sim, Sukin and Veis, Libor and Aspuru-Guzik, Alán},
title = {{Quantum Chemistry in the Age of Quantum Computing}},
journal = {Chemical Reviews},
volume = {119},
number = {19},
pages = {10856-10915},
year = {2019},
doi = {10.1021/acs.chemrev.8b00803},
note ={PMID: 31469277},
URL = {https://doi.org/10.1021/acs.chemrev.8b00803},
}

@inproceedings{NISQ1,
author = {Kim, Junpyo and Min, Dongmoon and Cho, Jungmin and Jeong, Hyeonseong and Byun, Ilkwon and Choi, Junhyuk and Hong, Juwon and Kim, Jangwoo},
title = {{A Fault-Tolerant Million Qubit-Scale Distributed Quantum Computer}},
year = {2024},
publisher = {ACM},
address = {NY, USA},
url = {https://doi.org/10.1145/3620665.3640388},
doi = {10.1145/3620665.3640388},
pages = {1–19},
numpages = {19},
booktitle = {ACM International Conference on Architectural Support for Programming Languages and Operating Systems},
series = {ASPLOS '24}
}

@article{NISQ2,
  doi = {10.22331/q-2018-08-06-79},
  url = {https://doi.org/10.22331/q-2018-08-06-79},
  title = {{Quantum Computing in the NISQ Era and Beyond}},
  author = {Preskill, John},
  journal = {{Quantum}},
  issn = {2521-327X},
  publisher = {{Verein zur F{\"{o}}rderung des Open Access Publizierens in den Quantenwissenschaften}},
  volume = {2},
  pages = {79},
  month = aug,
  year = {2018}
}

@article{rus1,
author = {Wiebe, Nathan and Roetteler, Martin},
title = {{Quantum Arithmetic and Numerical Analysis using Repeat-until-Success Circuits}},
year = {2016},
issue_date = {January 2016},
publisher = {Rinton Press, Incorporated},
address = {Paramus, NJ},
volume = {16},
number = {1–2},
issn = {1533-7146},
journal = {Quantum Info. Comput.},
month = jan,
pages = {134–178},
numpages = {45}
}

@article{rus2,
  title={{Realization of a Quantum Neural Network using Repeat-until-Success Circuits in a Superconducting Quantum Processor}},
  author={Miguel Moreira and Gian Giacomo Guerreschi and Wouter Vlothuizen and J. F. Marques and J. van Straten and Shavindra P. Premaratne and X. Zou and H. Ali and Nandini Muthusubramanian and C. Zachariadis and J. van Someren and M. Beekman and N. Haider and Alessandro Bruno and Carmen Garcia Almudever and Anne Y. Matsuura and Leonardo DiCarlo},
  journal={npj Quantum Information},
  year={2022},
  volume={9},
  pages={1-7},
  url={https://api.semanticscholar.org/CorpusID:254926424}
}

@inproceedings{qrmap,
author = {Kim, Hyungseok and Jang, Enhyeok and Choi, Seungwoo and Kim, Youngmin and Ro, Won Woo},
title = {{QR-Map: A Map-Based Approach to Quantum Circuit Abstraction for Qubit Reuse Optimization}},
year = {2025},
isbn = {9798400712616},
publisher = {ACM},
address = {NY, USA},
url = {https://doi.org/10.1145/3695053.3731020},
doi = {10.1145/3695053.3731020},
booktitle = {Proceedings of the 52nd Annual International Symposium on Computer Architecture},
pages = {1568–1582},
numpages = {15},
location = {
},
series = {ISCA '25}
}

@inproceedings{caqr,
author = {Hua, Fei and Jin, Yuwei and Chen, Yanhao and Vittal, Suhas and Krsulich, Kevin and Bishop, Lev S. and Lapeyre, John and Javadi-Abhari, Ali and Zhang, Eddy Z.},
title = {{CaQR: A Compiler-Assisted Approach for Qubit Reuse through Dynamic Circuit}},
year = {2023},
isbn = {9781450399180},
publisher = {ACM},
address = {NY, USA},
url = {https://doi.org/10.1145/3582016.3582030},
doi = {10.1145/3582016.3582030},
booktitle = {Proceedings of the 28th ACM International Conference on Architectural Support for Programming Languages and Operating Systems, Volume 3},
pages = {59–71},
numpages = {13},
location = {Vancouver, BC, Canada},
series = {ASPLOS 2023}
}

@INPROCEEDINGS {qce_reuse,
author = { Brandhofer, Sebastian and Polian, Ilia and Krsulich, Kevin },
booktitle = { 2023 IEEE International Conference on Quantum Computing and Engineering (QCE) },
title = {{ Optimal Qubit Reuse for Near-Term Quantum Computers }},
year = {2023},
volume = {},
ISSN = {},
pages = {859-869},
doi = {10.1109/QCE57702.2023.00100},
url = {https://doi.ieeecomputersociety.org/10.1109/QCE57702.2023.00100},
publisher = {IEEE Computer Society},
address = {Los Alamitos, CA, USA},
month =sep}

@article{nc_mcm_measure,
author = {Hothem, Daniel and Hines, Jordan and Baldwin, Charles and Gresh, Dan and Blume-Kohout, Robin and Proctor, Timothy},
year = {2025},
month = {07},
pages = {},
title = {{Measuring Error Rates of Mid-circuit Measurements}},
volume = {16},
journal = {Nature Communications},
doi = {10.1038/s41467-025-60923-x}
}

@article{qasmbench,
author = {Li, Ang and Stein, Samuel and Krishnamoorthy, Sriram and Ang, James},
title = {{QASMBench: A Low-Level Quantum Benchmark Suite for NISQ Evaluation and Simulation}},
year = {2023},
issue_date = {June 2023},
publisher = {ACM},
address = {NY, USA},
volume = {4},
number = {2},
url = {https://doi.org/10.1145/3550488},
doi = {10.1145/3550488},
journal = {ACM Transactions on Quantum Computing},
month = feb,
articleno = {10},
numpages = {26}
}

@misc{Web:IBM,
	author = {{IBM}},
	title = {{IBM Quantum Computing}},
    year = {2025},
    note = {Accessed: Sep 24th 2025. [Online]. Available: \url{https://www.ibm.com/quantum}}
}

@misc{Web:qiskit_tanspiler,
	author = {{IBM}},
	title = "{{IBM Qiskit Transpiler}}",
    year   = {2025},
    note   = {Accessed: Oct 3rd 2025. [Online]. Available: \url{https://quantum.cloud.ibm.com/docs/en/api/qiskit/transpiler}}
}

@misc{fidelity,
      title={{Quantum Computing with Qiskit}}, 
      author={Ali Javadi-Abhari and Matthew Treinish and Kevin Krsulich and Christopher J. Wood and Jake Lishman and Julien Gacon and Simon Martiel and Paul D. Nation and Lev S. Bishop and Andrew W. Cross and Blake R. Johnson and Jay M. Gambetta},
      year={2024},
      eprint={2405.08810},
      archivePrefix={arXiv},
      primaryClass={quant-ph},
      url={https://arxiv.org/abs/2405.08810}, 
}

@inproceedings{ScaffCC,
author = {Javadi-Abhari, Ali and Patil, Shruti and Kudrow, Daniel and Heckey, Jeff and Lvov, Alexey and Chong, Frederic T. and Martonosi, Margaret},
title = {{ScaffCC: a Framework for Compilation and Analysis of Quantum Computing Programs}},
year = {2014},
isbn = {9781450328708},
publisher = {ACM},
address = {NY, USA},
url = {https://doi.org/10.1145/2597917.2597939},
doi = {10.1145/2597917.2597939},
booktitle = {Proceedings of the 11th ACM Conference on Computing Frontiers},
articleno = {1},
numpages = {10},
location = {Cagliari, Italy},
series = {CF '14}
}

@article{fermihedral,
  author = {Liu, Yuhao and Che, Shize and Zhou, Junyu and Shi, Yunong and Li, Gushu},
  title = {Fermihedral: On the Optimal Compilation for Fermion-to-Qubit Encoding},
  year = {2024},
  isbn = {979-8-4007-0386},
  pages = {382–397},
  volume = {3},
  number = {1},
  publisher = {ACM},
  url = {https://doi.org/10.1145/3620666.3651371},
  doi = {10.1145/3620666.3651371},
  journal = {ASPLOS},
  address = {La Jolla, CA, USA},
  series = {ASPLOS '24},
}

@inproceedings{Paulihedral,
author = {Li, Gushu and Wu, Anbang and Shi, Yunong and Javadi-Abhari, Ali and Ding, Yufei and Xie, Yuan},
title = {{Paulihedral: a Generalized Block-Wise Compiler Optimization Framework for Quantum Simulation Kernels}},
year = {2022},
isbn = {9781450392051},
publisher = {ACM},
address = {NY, USA},
booktitle = {Proceedings of the 27th ACM International Conference on Architectural Support for Programming Languages and Operating Systems},
pages = {554–569},
numpages = {16},
location = {Lausanne, Switzerland},
series = {ASPLOS '22}
}

@inproceedings{SQUARE,
author = {Ding, Yongshan and Wu, Xin-Chuan and Holmes, Adam and Wiseth, Ash and Franklin, Diana and Martonosi, Margaret and Chong, Frederic T.},
title = {{SQUARE: Strategic Quantum Ancilla Reuse for Modular Quantum Programs via Cost-Effective Uncomputation}},
year = {2020},
isbn = {9781728146614},
publisher = {IEEE Press},
url = {https://doi.org/10.1109/ISCA45697.2020.00054},
doi = {10.1109/ISCA45697.2020.00054},
booktitle = {Proceedings of the ACM/IEEE 47th Annual International Symposium on Computer Architecture},
pages = {570–583},
numpages = {14},
address = {Virtual Event},
series = {ISCA '20}
}

@INPROCEEDINGS{dyrem,
  author={Zhou, Kaiwen and Lu, Liqiang and Zhang, Hanyu and Xiang, Debin and Tao, Chenning and Zhao, Xinkui and Zheng, Size and Yin, Jianwei},
  booktitle={2025 62nd ACM/IEEE Design Automation Conference (DAC)}, 
  title={{DyREM: Dynamically Mitigating Quantum Readout Error with Embedded Accelerator}}, 
  year={2025},
  volume={},
  number={},
  pages={1-7},
  doi={10.1109/DAC63849.2025.11132635},
  publisher = {IEEE Press},
  address = {CA, USA},
}

@inproceedings{sprem,
author = {Zhang, Hanyu and Lu, Liqiang and Tan, Siwei and Zheng, Size and Yu, Jia and Yin, Jianwei},
title = {{SpREM: Exploiting Hamming Sparsity for Fast Quantum Readout Error Mitigation}},
year = {2024},
isbn = {9798400706011},
publisher = {ACM},
address = {NY, USA},
url = {https://doi.org/10.1145/3649329.3655675},
doi = {10.1145/3649329.3655675},
booktitle = {Proceedings of the 61st ACM/IEEE Design Automation Conference},
articleno = {91},
numpages = {6},
location = {CA, USA},
series = {DAC '24}
}

@article{mcm_random,
author = {Govia, L and Jurcevic, P and Wood, Christopher and Kanazawa, N and Merkel, S and McKay, D},
year = {2023},
month = {12},
pages = {},
title = {{A Randomized Benchmarking Suite for Mid-Circuit Measurements}},
volume = {25},
journal = {New Journal of Physics}
}

@article{BV_algo,
author = {Bernstein, Ethan and Vazirani, Umesh},
title = {Quantum Complexity Theory},
journal = {SIAM Journal on Computing},
volume = {26},
number = {5},
pages = {1411-1473},
year = {1997},
URL = {https://doi.org/10.1137/S0097539796300921}
}

@misc{Web:google_quantum,
	author = {{Google}},
	title = {{Google Quantum Computer}},
    year = {2025},
    note = {Accessed: Sep 16th 2025. [Online]. Available: \url{https://quantumai.google/quantumcomputer}}
}

@misc{koh2025readouterrormitigationmidcircuit,
      title={{Readout Error Mitigation for Mid-Circuit Measurements and Feedforward}}, 
      author={Jin Ming Koh and Dax Enshan Koh and Jayne Thompson},
      year={2025},
      eprint={2406.07611},
      archivePrefix={arXiv},
      primaryClass={quant-ph},
      url={https://arxiv.org/abs/2406.07611}, 
}

@article{nc_leakage,
    title = {{Removing Leakage-induced Correlated Errors in Superconducting Quantum Error Correction}},
    author	= {Matt McEwen and Dvir Kafri and Jimmy Chen and Juan Atalaya and Kevin Satzinger and Chris Quintana and Paul Victor Klimov and Daniel Sank and Craig Michael Gidney and Austin Fowler and Frank Carlton Arute and Kunal Arya and Bob Benjamin Buckley and Brian Burkett and Nicholas Bushnell and Benjamin Chiaro and Roberto Collins and Sean Demura and Andrew Dunsworth and Catherine Erickson and Brooks Riley Foxen and Marissa Giustina and Trent Huang and Sabrina Hong and Evan Jeffrey and Seon Kim and Kostyantyn Kechedzhi and Fedor Kostritsa and Pavel Laptev and Anthony Megrant and Xiao Mi and Josh Mutus and Ofer Naaman and Matthew Neeley and Charles Neill and Murphy Yuezhen Niu and Alexandru Paler and Nick Redd and Pedram Roushan and Ted White and Jamie Yao and Ping Yeh and Adam Jozef Zalcman and Yu Chen and Vadim Smelyanskiy and John Martinis and Hartmut Neven and J. Kelly and Alexander Korotkov and Andre Gregory Petukhov and Rami Barends},
    journal = {Nature Communications},
  volume = {12},
  year = {2021},
  pages={1761},
  publisher = {Springer Science and Business Media LLC},

}

@inproceedings{CADD,
  author={Seif Alireza and Liao Haoran and Tripathi Vinay and Krsulich Kevin and Malekakhlagh Moein and Amico Mirko and Jurcevic Petar and Javadi-Abhari Ali},
  booktitle={2024 ACM/IEEE 51st Annual International Symposium on Computer Architecture (ISCA)}, 
  title={{Suppressing Correlated Noise in Quantum Computers via Context-Aware Compiling}}, 
  year={2024},
  volume={},
  number={},
  pages={310-324},
  publisher = {IEEE Press},
  address = {Buenos Aires, Argentina},
  doi={10.1109/ISCA59077.2024.00031}
}

@ARTICLE{quantum_finance,
  author={Egger, Daniel J. and Gambella, Claudio and Marecek, Jakub and McFaddin, Scott and Mevissen, Martin and Raymond, Rudy and Simonetto, Andrea and Woerner, Stefan and Yndurain, Elena},
  journal={IEEE Transactions on Quantum Engineering}, 
  title={Quantum Computing for Finance: State-of-the-Art and Future Prospects}, 
  year={2020},
  volume={1},
  number={},
  pages={1-24},
  doi={10.1109/TQE.2020.3030314}
}

@article{readout_error,
  title = {Efficient quantum readout-error mitigation for sparse measurement outcomes of near-term quantum devices},
  author = {Yang, Bo and Raymond, Rudy and Uno, Shumpei},
  journal = {Phys. Rev. A},
  volume = {106},
  issue = {1},
  pages = {012423},
  numpages = {14},
  year = {2022},
  month = {Jul},
  publisher = {American Physical Society},
  doi = {10.1103/PhysRevA.106.012423},
  url = {https://link.aps.org/doi/10.1103/PhysRevA.106.012423}
}

@INPROCEEDINGS{9773181,
  author={Ueno, Yosuke and Kondo, Masaaki and Tanaka, Masamitsu and Suzuki, Yasunari and Tabuchi, Yutaka},
  booktitle={2022 IEEE International Symposium on High-Performance Computer Architecture (HPCA)}, 
  title={QULATIS: A Quantum Error Correction Methodology toward Lattice Surgery}, 
  year={2022},
  volume={},
  number={},
  pages={274-287},
  publisher = {IEEE Press},
  address = {Seoul, South Korea},
  doi={10.1109/HPCA53966.2022.00028}
}

@inbook{DAC_error_correction,
author = {Ueno, Yosuke and Kondo, Masaaki and Tanaka, Masamitsu and Suzuki, Yasunari and Tabuchi, Yutaka},
title = {QECOOL: On-Line Quantum Error Correction with a Superconducting Decoder for Surface Code},
year = {2022},
isbn = {9781665432740},
publisher = {IEEE Press},
url = {https://doi.org/10.1109/DAC18074.2021.9586326},
booktitle = {Proceedings of the 58th Annual ACM/IEEE Design Automation Conference},
pages = {451–456},
address = {CA, USA},
numpages = {6}
}

@INPROCEEDINGS {TILT,
author = { Wu, Xin-Chuan and Debroy, Dripto M. and Ding, Yongshan and Baker, Jonathan M. and Alexeev, Yuri and Brown, Kenneth R. and Chong, Frederic T. },
booktitle = { 2021 IEEE International Symposium on High-Performance Computer Architecture (HPCA) },
title = {{ TILT: Achieving Higher Fidelity on a Trapped-Ion Linear-Tape Quantum Computing Architecture }},
year = {2021},
volume = {},
ISSN = {},
pages = {153-166},
doi = {10.1109/HPCA51647.2021.00023},
url = {https://doi.ieeecomputersociety.org/10.1109/HPCA51647.2021.00023},
publisher = {IEEE Computer Society},
address = {CA, USA},
month =mar}

@inproceedings{chen_error,
author = {Chu, Cheng and Chia, Nai-Hui and Jiang, Lei and Chen, Fan},
title = {QMLP: An Error-Tolerant Nonlinear Quantum MLP Architecture using Parameterized Two-Qubit Gates},
year = {2022},
isbn = {9781450393546},
publisher = {ACM},
address = {NY, USA},
url = {https://doi.org/10.1145/3531437.3539719},
doi = {10.1145/3531437.3539719},
booktitle = {Proceedings of the ACM/IEEE International Symposium on Low Power Electronics and Design},
articleno = {4},
numpages = {6},
location = {Boston, MA, USA},
series = {ISLPED '22}
}
